# Coupling of Optical Lumped Nanocircuit Elements and Effects of Substrates

Andrea Alù, Alessandro Salandrino, and Nader Engheta[*]

University of Pennsylvania

Department of Electrical and Systems Engineering

Philadelphia, Pennsylvania 19104, U.S.A.

**Abstract**

We present here a model for the coupling among small nanoparticles excited by an optical electric field in the framework of our optical lumped nanocircuit theory [N. Engheta, A. Salandrino, and A. Alù, *Phys. Rev. Lett.* **95**, 095504 (2005)]. We derive how this coupling affects the corresponding nanocircuit model by adding controlled sources that depend on the optical voltages applied on the coupled particles. With the same technique, we can model also the presence of a substrate underneath nanocircuit elements, relating its presence to the coupling with a properly modeled image nanoparticle. These results are of importance in the understanding and the design of complex optical nanocircuits at infrared and optical frequencies.

*1.   Introduction*

In a recent paper [1] we have proposed a paradigm for extending the classic circuit concepts, commonly available at microwave and lower frequencies, to higher frequencies and in particular to the optical domain. To this end, we have shown how a proper combination of plasmonic and non-plasmonic nanoparticles may allow envisioning the design of a complex nanocircuit at frequencies where the conventional lumped circuit elements, like lumped capacitors, inductors and resistors, had not been considered before. Following our work in [1] for an isolated nanoparticle, we have shown how these concepts may be applied to the design of relatively complex functional circuits, i.e., planar nanotransmission lines [2], nanowires of linear chain of particles [3] and 3-D nanotransmission line metamaterials [4]. Moreover, in [5]-[6] we have analyzed in details how a pair of touching nanocircuit elements may be indeed envisioned as in the series or the parallel configuration depending on their orientation with respect to the impressed optical electric field. All these concepts are important steps towards the possibility of synthesizing a complex optical *nanocircuit board* with the functionalities analogous to a classic microwave circuit (e.g., filtering, waveguiding, multiplexing…). In the following, we provide an extended circuit model for the general case of a pair of arbitrary particles placed in proximity of each other, in order to analyze the interactions between such nanocircuit elements in this more general case. Such concepts may then be extended to more arbitrary configurations of nanoparticles, and may provide methods to model a planar substrate underneath these nanocircuits. Ways of avoiding unwanted coupling among nanoparticles are suggested following the present analysis. These results may allow the modeling of a properly designed collection of nanoparticles closer to a real circuit design, which is of interest for a large number of potential applications in applied optics. An $e^{-i\omega t}$ is considered throughout this manuscript.

## 2. *An Isolated Nanocircuit Element*

Following the results of [1], an isolated nanoparticle illuminated by a uniform electric field $\mathbf{E}_0$ may be regarded as a lumped nanocircuit element with complex impedance $Z_{nano}$, as depicted schematically in Fig. 1.

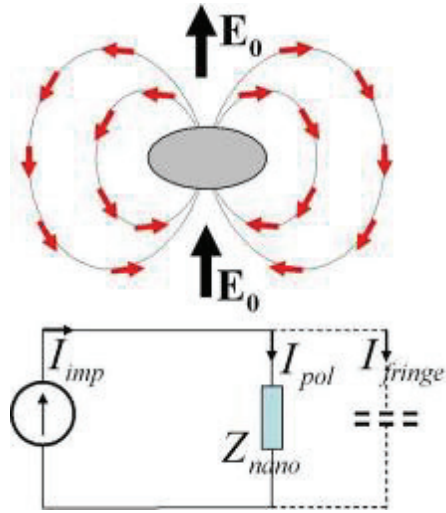

Fig. 1. (Color online). A nanoparticle illuminated by a uniform optical electric field $\mathbf{E}_0$ (black arrows) may be viewed in terms of the circuit analogy presented in [1] as a lumped impedance $Z_{nano}$ excited by the impressed current generator $I_{imp}$ and loaded with the fringe capacitance associated with its fringe dipolar fields (red arrows).

Such nanoimpedance $Z_{nano}$ may be defined, analogously to the classic concept of impedance in circuit theory, as the ratio of the optical voltage $V$ across the "ends" (or the "terminals") of the nanoelement and the displacement current $I_{pol}$ circulating across it. Such impedance is a fixed quantity, depending only on the geometry of the particle and its constituent materials and possibly on the orientation of the applied field in case of particles with anisotropic polarizability. In the nanocircuit paradigm [1] the voltage $V$, defined as the averaged potential difference between the "top" and the "bottom" of the particle, and the displacement current $I_{pol}$, defined as the integral flux of the displacement current $\mathbf{J}_d = -i\omega\mathbf{D}$ flowing across the top or bottom "terminal" of the nanoparticle, where $\mathbf{D}$ is the local electric displacement vector, satisfy Kirchhoff's circuit laws, and therefore such definitions justify a complete correspondence and physical analogy between the scattering problem of an isolated particle (Fig. 1 top) and its equivalent circuit representation in terms of its impedance $Z_{nano}$ (Fig. 1 bottom). In particular, the nanocircuit is completed by an impressed current generator of amplitude $I_{imp}$, associated with the external exciting field, and a parallel fringe capacitance of impedance $Z_{fringe}$, related to the dipolar fields around the nanoparticle. For a spherical nanoparticle, these expressions take the form [1]:

$$\begin{aligned} Z_{nano} &= \left(-i\omega\varepsilon\pi R\right)^{-1} \\ I_{imp} &= -i\omega\left(\varepsilon-\varepsilon_0\right)\pi R^2 \left|\mathbf{E}_0\right|, \\ Z_{fringe} &= \left(-i\omega 2\pi R\varepsilon_0\right)^{-1} \end{aligned} \quad (1)$$

where $\varepsilon$ is the nanosphere's material permittivity, $R$ its radius and $\varepsilon_0$ the background permittivity.

This description is consistent with the circuit sketch of Fig. 1, and it implies that a non-plasmonic (plasmonic) isolated nanoparticle may act as a lumped nanocapacitance (nanoinductance), due to the positive (negative) sign of the real part of its permittivity (see eq. (1)). In this analogy, the presence of material loss corresponds to a nanoresistor.

This situation may be generalized by altering the spherical symmetry of the particle shape. For instance, if one considers an ellipsoidal particle with semi-axes $a$, $b$, and $c$ (where $a > b > c$) and permittivity $\varepsilon$, following similar steps as in the spherical case, the effective impedance of such ellipsoidal nanoparticle excited by a uniform electric field parallel to the $c$ axis becomes:

$$Z_{nano} = \frac{U(a,b,c)}{-i\omega\varepsilon\pi ab}, \qquad (2)$$

with $U(a,b,c) \equiv \left[ \dfrac{2c \int_{S_{up}} \sqrt{(c^2+\eta)(c^2+\zeta)}dS}{\sqrt{(a^2-c^2)(b^2-c^2)}S_{up}} \right]$, where $\eta, \zeta$ are the transverse coordinates on the ellipsoid in a confocal ellipsoidal reference system and the integral is taken over half of the ellipsoidal surface with area $S_{up}$ [7].

The fringe impedance and the impressed current are given in this case by the expressions:

$$Z_{fringe} = \frac{U(a,b,c)}{-i\omega\varepsilon_0 \pi ab \dfrac{1-L_z(0)}{L_z(0)}}, \qquad (3)$$

$$I_{imp} = -i\omega(\varepsilon - \varepsilon_0)\pi ab E_0$$

where $L_z(0) \equiv \dfrac{abc}{2} \int_0^\infty \dfrac{dq}{(c^2+q)\sqrt{(q+a^2)(q+b^2)(q+c^2)}}$. Such expressions converge to Eqs. (1) when $a = b = c$, i.e., in the spherical geometry. Equivalently, the effective circuit elements for a generic nanoellipsoid, i.e., its equivalent capacitance $C_{eq,ellipsoid}$, inductance $L_{eq,ellipsoid}$, resistance $R_{eq,ellipsoid}$ and the equivalent fringe capacitance $C_{eq,fringe}$ are given by:

$$C_{eq,ellipsoid} = \frac{\pi ab \, \text{Im}(\varepsilon)}{U(a,b,c)} \quad \text{if } \text{Re}(\varepsilon) > 0 \tag{4}$$

$$L_{eq,ellipsoid} = \frac{U(a,b,c)}{-\omega^2 \pi ab \, \text{Re}(\varepsilon)} \quad \text{if } \text{Re}(\varepsilon) < 0 \tag{5}$$

$$G_{eq,ellipsoid} = \frac{\omega \pi ab \, \text{Im}(\varepsilon)}{U(a,b,c)} \tag{6}$$

$$C_{eq,fringe} = \frac{\varepsilon_o \pi ab \frac{1-L_z(0)}{L_z(0)}}{U(a,b,c)}. \tag{7}$$

Varying the orientation of the exciting electric field or the aspect ratio of the ellipsoid one can modify the effective impedance associated with the nanoparticle, effectively adding new degrees of freedom to the possibility of synthesizing the desired circuit response with an isolated dielectric or plasmonic nanoparticle.

The impedance $Z_{nano}$ enters into resonance with $Z_{fringe}$ in the circuit of Fig. 1 when $Z_{fringe} = -Z_{nano}$, i.e., for:

$$\varepsilon = -\frac{1-L_z(0)}{L_z(0)} \varepsilon_0, \tag{8}$$

which coincides with the "quasi-static" condition for the resonant scattering of an ellipsoid [8]. This collapses to the usual $\varepsilon = -2\varepsilon_0$ in the spherical case. Such properties confirm the equivalence between the scattering phenomenon depicted in Fig. 1 (top) and the equivalent circuit of Fig. 1 (bottom).

The three basic elements of any linear circuit, R, L, and C, which are at the core of many complex circuits, may therefore be considered available at infrared and optical frequencies following this paradigm, since dielectric and plasmonic materials are naturally available in these frequency regimes [8]-[9].

What happens, however, when the nanoparticle is no longer an isolated system, but instead it is located in close proximity of other nanoparticles? For certain configurations (e.g., linear chains, periodic arrays and lattices) this problem has been solved analytically [2]-[4], showing how in those specific cases the circuit analogy still holds effectively, despite the fact that the fringing dipolar fields of all the particles do interact with each other. In [5]-[6], moreover, the rigorous analytical solution of a geometry consisting of two conjoined nanoparticles has been presented, showing how the nanocircuit analogy also holds in this relatively complex scenario. It should be underlined, however, that unlike the case of a classic circuit board, where the different lumped circuit elements are functionally isolated from the external world and interact with the other circuit elements on the board only through their terminals through which the *conduction* current flows in and out following specific paths, here the *displacement* current $\mathbf{J}_d = -i\omega \mathbf{D}$ may spread out in the surrounding space, resulting in coupling among different nanoelements, in a manner that may be undesirable. The main difference between the two scenarios is indeed related to the fact that the substrate or background material where the classic RF circuits are printed is usually poorly conductive in order to avoid unwanted leakage of conduction currents. In the optical nanocircuits, however, air or standard dielectrics where the nanocircuit elements may be printed or deposited have a permittivity, which plays the role of conductivity in our nanocircuit paradigm, often comparable in value with those of the nanocircuit elements. This may imply an unwanted leakage of the displacement current, with consequent coupling among different nanocircuit elements that by design should not be necessarily connected with each other for the correct functionalities of the nanocircuit. In the following we address these specific points and extend the nanocircuit analogy of Fig. 1, valid for an isolated nanoparticle, into more complex scenarios.

## *3. Interaction and Coupling between Two Nanocircuit Elements*

Extending the previous concepts to generic configurations with more than one nanoparticle, e.g., the case of two nanospheres with radii $R_1$ and $R_2$, permittivities $\varepsilon_1$ and $\varepsilon_2$, and with a center-to-center distance $d$, is of interest as a step towards the design of complex nanocircuits and understanding the mechanism of coupling among nanocircuit elements in the general case. A sketch of this configuration is shown in Fig. 2 (top).

In the case at hand, the fringing dipolar fields from each of the two nanoparticles interact and are generally modified by the presence of the other particle, altering the circuit representation for each of the two particles. In the first approximation, provided that the nanospheres are not extremely close and that we are far from higher-order resonances, we may consider their interaction as described by the induced dipoles in each of the particles. In the following paragraphs, we analytically show that these and similar configurations may be effectively treated as "coupled" nano-circuits, each representing one of the nanospheres (see Fig. 2 bottom). Each circuit in the figure includes the capacitive or inductive impedance of the given nanosphere, the capacitive impedance related to the fringe field, and the independent current source representing the impressed field on this sphere, as in Fig. 1. However, in addition, each circuit also needs to have a "dependent" current source representing the influence of the field of other particle(s) on this sphere. In other words, the interaction among the particles here may be exhibited by using such dependent sources. The value of each dependent current source in Fig. 2 may be explicitly derived in terms of the potential difference across the other nanosphere, in analogy with the expressions in the previous section, as we show in the following.

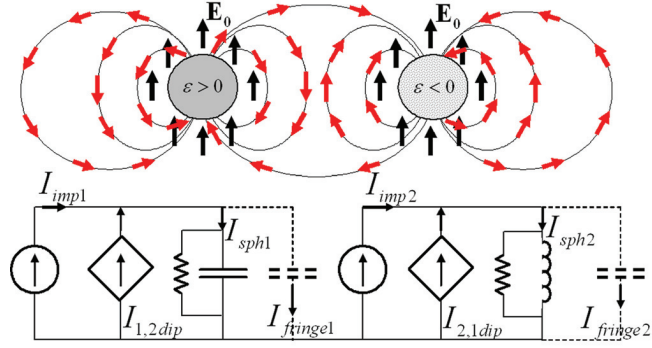

Fig. 2. (Color online). Following Fig. 1, a coupled nanocircuit in the optical domain, with optical field coupling between two adjacent nanospheres as two coupled nanoelements.

*a) Modeling the coupling between two nanospheres*

Consider the situation depicted in Fig. 2 (top), i.e., two nanospheres with radii $R_1$ and $R_2$, permittivities $\varepsilon_1$ and $\varepsilon_2$, and with a center-to-center distance $d$. In general the pair of spheres is excited by the uniform electric field $\mathbf{E}_0$ with arbitrary orientation. The electric field inside and outside the first sphere under the dipolar approximation is given by [10]:

$$\mathbf{E}_1^{in} = \frac{3\varepsilon_0}{\varepsilon_1 + 2\varepsilon_0}(\mathbf{E}_0 + \mathbf{E}_{12}) \tag{9}$$

$$\begin{cases} \mathbf{E}_1^{out} = \mathbf{E}_0 + \dfrac{3\dfrac{\mathbf{r}}{|\mathbf{r}|}\left(\mathbf{p}_1 \cdot \dfrac{\mathbf{r}}{|\mathbf{r}|}\right) - \mathbf{p}_1}{4\pi\varepsilon_0 |\mathbf{r}|^3} \\ \mathbf{p}_1 = 4\pi\varepsilon_0 \dfrac{\varepsilon_1 - \varepsilon_0}{\varepsilon_1 + 2\varepsilon_0} R_1^3 (\mathbf{E}_0 + \mathbf{E}_{12}) \end{cases} \tag{10}$$

In (9)-(10) $\mathbf{E}_{12}$ is the field originated from the second sphere, but evaluated at the center of the first sphere, which has been taken to coincide with the origin, and $\mathbf{r}$ is the observation vector.

Assuming for simplicity that the applied field is the same $\mathbf{E}_0$ at the second nanosphere, we get the following expressions for this second sphere:

$$\mathbf{E}_2^{in} = \frac{3\varepsilon_0}{\varepsilon_2 + 2\varepsilon_0}(\mathbf{E}_0 + \mathbf{E}_{21}) \tag{11}$$

$$\begin{cases} \mathbf{E}_2^{out} = \mathbf{E}_0 + \dfrac{3\dfrac{\mathbf{r}-\mathbf{d}}{|\mathbf{r}-\mathbf{d}|}\left(\mathbf{p}_2 \cdot \dfrac{\mathbf{r}-\mathbf{d}}{|\mathbf{r}-\mathbf{d}|}\right) - \mathbf{p}_2}{4\pi\varepsilon_0 |\mathbf{r}-\mathbf{d}|^3} \\ \mathbf{p}_2 = 4\pi\varepsilon_0 \dfrac{\varepsilon_2 - \varepsilon_0}{\varepsilon_2 + \varepsilon_0} R_2^3 (\mathbf{E}_0 + \mathbf{E}_{21}) \end{cases} \tag{12}$$

$\mathbf{E}_{21}$ is the field emitted from the first sphere, evaluated at the center of the second sphere and $\mathbf{d}$ is the vector that locate the generic position of the second sphere.

Since $\mathbf{E}_{12} = \mathbf{E}_2^{out}\big|_{\mathbf{r}=0}$ and $\mathbf{E}_{21} = \mathbf{E}_1^{out}\big|_{\mathbf{r}=\mathbf{d}}$ we may evaluate their expression in closed form using (10), (12), obtaining:

$$\begin{pmatrix} \mathbf{E}_{12} \\ \mathbf{E}_{21} \end{pmatrix} = \begin{pmatrix} |\mathbf{d}|^3 \overline{\mathbf{I}} & -\gamma_2(3\overline{\mathbf{D}} - \overline{\mathbf{I}}) \\ \gamma_1(3\overline{\mathbf{D}} - \overline{\mathbf{I}}) & -|\mathbf{d}|^3 \overline{\mathbf{I}} \end{pmatrix}^{-1} \cdot \begin{pmatrix} \gamma_2(3\overline{\mathbf{D}} - \overline{\mathbf{I}}) \cdot \mathbf{E}_0 \\ -\gamma_1(3\overline{\mathbf{D}} - \overline{\mathbf{I}}) \cdot \mathbf{E}_0 \end{pmatrix}, \tag{13}$$

with: $\overline{\mathbf{D}} = \dfrac{\mathbf{d}}{|\mathbf{d}|}\dfrac{\mathbf{d}}{|\mathbf{d}|}$, $\overline{\mathbf{I}} = \begin{pmatrix} 1 & 0 & 0 \\ 0 & 1 & 0 \\ 0 & 0 & 1 \end{pmatrix}$, and $\gamma_i = \dfrac{\varepsilon_i - \varepsilon_0}{\varepsilon_i + 2\varepsilon_0} R_i^3$.

Eq. (13) is completely general, for an arbitrary orientation of $\mathbf{E}_0$ and $\mathbf{d}$. If $\mathbf{E}_0 \parallel \mathbf{d}$, we will get the simplified expression:

$$\begin{cases} \mathbf{E}_{12} = \dfrac{2\gamma_2\left(d^3+2\gamma_1\right)}{d^6-4\gamma_1\gamma_2}\mathbf{E_0} \\ \mathbf{E}_{21} = \dfrac{2\gamma_1\left(d^3+2\gamma_2\right)}{d^6-4\gamma_1\gamma_2}\mathbf{E_0} \end{cases} \quad (14)$$

whereas when $\mathbf{E_0} \perp \mathbf{d}$ we obtain:

$$\begin{cases} \mathbf{E}_{12} = \dfrac{\gamma_2\left(\gamma_1-d^3\right)}{d^6-\gamma_1\gamma_2}\mathbf{E_0} \\ \mathbf{E}_{21} = \dfrac{\gamma_1\left(\gamma_2-d^3\right)}{d^6-\gamma_1\gamma_2}\mathbf{E_0} \end{cases} \quad (15)$$

In these two specific cases of interest, as evident from (14) and (15), $\mathbf{E}_{12} \parallel \mathbf{E}_{21} \parallel \mathbf{E_0}$, implying that in each of the two circuits representing the two nanoparticles the coupling effect simply adds an extra term to the equivalent impressed current. This is depicted in Fig. 2, where the extra current generators have amplitudes:

$$\begin{aligned} I_{12,dip} &= -i\omega\pi\left|\mathbf{E}_{12}\right|\left(\varepsilon_1-\varepsilon_0\right)R_1^2 \\ I_{21,dip} &= -i\omega\pi\left|\mathbf{E}_{21}\right|\left(\varepsilon_2-\varepsilon_0\right)R_2^2 \end{aligned} \quad (16)$$

which satisfy reciprocity.

We note that the expressions of $\mathbf{E}_{21}$ and $\mathbf{E}_{12}$, which have been derived in closed form in (15) by taking into account the overall coupling between the nanoparticles, depend on the geometry of both nanoparticles, and therefore the two resulting circuits are expectedly coupled with each other. In particular, the magnitude of the extra dependent current exciting the first particle, $I_{12,dip}$, is directly proportional to the magnitude of the induced dipole moment $\mathbf{p}_2$ on the second particle, which may be related to the averaged potential difference $\langle V \rangle_{sphere\,2}$ across the equivalent impedance of the sphere 2 as:

$|\mathbf{p}_2| = 4\pi\varepsilon_0 R_2^2 \langle V \rangle_{sphere\,2}$. Substituting this into the expression for $I_{12,dep.}$, it becomes evident that $I_{12,dep.}$ depends on $\langle V \rangle_{sphere\,2}$, and hence it is a dependent current source for sphere 1. By analogy, similar considerations for $I_{21,dep.}$ for sphere 2 hold, showing that effectively the extra current generators are dependent generators, functions of the voltage applied on the coupled nanoelements.

*b) Modeling the coupling between two ellipsoids*

The previous analysis may be generalized to the case of two ellipsoids, under the assumption that the impressed optical electric field $\mathbf{E}_0$ is directed along one of their principal axes. Following similar steps, and assuming that the center-to-center vector $\mathbf{d}$ is along the semiaxis $c$ of the ellipsoids, we get the following expressions in the case of $\mathbf{E}_0 \perp \mathbf{d}$:

$$\mathbf{E}_{12} = \frac{\dfrac{(\varepsilon_0 - \varepsilon_1)L_1(d^2 - c_1^2)}{1 + (\varepsilon_1 - \varepsilon_0)L_1(0)}\left[1 + \dfrac{(\varepsilon_0 - \varepsilon_2)L_2(d^2 - c_2^2)}{1 + (\varepsilon_2 - \varepsilon_0)L_2(0)}\right]}{1 - \dfrac{(\varepsilon_0 - \varepsilon_1)L_1(d^2 - c_1^2)}{1 + (\varepsilon_1 - \varepsilon_0)L_1(0)}\dfrac{(\varepsilon_0 - \varepsilon_2)L_2(d^2 - c_2^2)}{1 + (\varepsilon_2 - \varepsilon_0)L_2(0)}}\mathbf{E}_0 \quad , \qquad (17)$$

$$\mathbf{E}_{21} = \frac{\dfrac{(\varepsilon_0 - \varepsilon_2)L_2(d^2 - c_2^2)}{1 + (\varepsilon_2 - \varepsilon_0)L_2(0)}\left[1 + \dfrac{(\varepsilon_0 - \varepsilon_1)L_1(d^2 - c_1^2)}{1 + (\varepsilon_1 - \varepsilon_0)L_1(0)}\right]}{1 - \dfrac{(\varepsilon_0 - \varepsilon_1)L_1(d^2 - c_1^2)}{1 + (\varepsilon_1 - \varepsilon_0)L_1(0)}\dfrac{(\varepsilon_0 - \varepsilon_2)L_2(d^2 - c_2^2)}{1 + (\varepsilon_2 - \varepsilon_0)L_2(0)}}\mathbf{E}_0$$

where $L_i(\xi_i) = \dfrac{a_i b c_i}{2}\int_{\xi_i}^{\infty}\dfrac{dq}{(b^2 + q)f_i(q)}$, and the two ellipsoids have semiaxes $a_1$, $b$, and $c_1$ (where $a_1 > b > c_1$) and $a_2$, $b$, and $c_2$ (where $a_2 > b > c_2$).

For the case of $\mathbf{E}_0 \parallel \mathbf{d}$ we get analogously:

$$\mathbf{E}_{12} = \frac{A_1(1+A_2)}{1-A_1 A_2}\mathbf{E}_0$$

$$\mathbf{E}_{21} = \frac{A_2(1+A_1)}{1-A_1 A_2}\mathbf{E}_0 \quad , \tag{18}$$

with

$$\begin{cases} A_1 = \dfrac{(\varepsilon_0 - \varepsilon_1)L_1(d^2 - c_1^2)}{1+(\varepsilon_1 - \varepsilon_0)L_1(0)} + \dfrac{\varepsilon_1 - \varepsilon_0}{1+(\varepsilon_1 - 1)L_1(0)}\dfrac{a_1 b c_1}{d\sqrt{(c_1^2 - a_1^2 - d^2)(c_1^2 - b^2 - d^2)}} \\ A_2 = \dfrac{(\varepsilon_0 - \varepsilon_2)L_2(d^2 - c_2^2)}{1+(\varepsilon_2 - \varepsilon_0)L_2(0)} + \dfrac{(\varepsilon_2 - \varepsilon_0)}{1+(\varepsilon_2 - \varepsilon_0)L_2(0)}\dfrac{a_2 b c_2}{d\sqrt{(c_2^2 - a_2^2 - d^2)(c_2^2 - b^2 - d^2)}} \end{cases}.$$

These expressions allows modeling of the coupling between the two ellipsoids with the circuit analogy of Fig. 2 and the formulas (2)-(7) derived in the previous section. The controlled generators in this case have expressions:

$$I_{12,dip} = -i\omega\pi|\mathbf{E}_{12}|(\varepsilon_1 - \varepsilon_0)a_1 c_1$$
$$I_{21,dip} = -i\omega\pi|\mathbf{E}_{21}|(\varepsilon_2 - \varepsilon_0)a_2 c_2 \quad , \tag{19}$$

where it is assumed that the polarization of the electric field is directed along the $b$ axis. The two configurations analyzed above, for which the impinging electric field is parallel or orthogonal to the vector $\mathbf{d}$, may be associated with the "quasi-series" or "quasi-parallel" interconnections among the particles. As shown in [5]-[6], when two nanoparticles are conjoined to each other with common interfaces normal to the orientation of the electric field, their equivalent nanoimpedances may be regarded as connected in the series configuration (since the effective displacement current that flows through them is effectively the same), whereas when their common interfaces are parallel to the electric field they may be regarded as in the parallel configuration (since the potential drop at their terminals is effectively the same). In this more general case here, the coupling between the nanoparticles cannot be considered strictly in parallel (series) when $\mathbf{E}_0 \parallel \mathbf{d}$ ($\mathbf{E}_0 \perp \mathbf{d}$) because some leakage of voltage (current) is indeed present due to the finite spacing between them. Since the total dimension of the nanocircuit is small compared with the wavelength in the background material, this leakage may, however, be considered small for some applications, and these quasi-series and quasi-parallel interconnections may even provide further degrees of freedom in the nanocircuit design. For this purpose, Fig. 3 shows the potential distribution for a "quasi-parallel" (Fig. 3a) and a "quasi-series" (Fig. 3b) configuration of two ellipsoids. It can be seen how in the quasi-parallel configuration of Fig. 3a, with the impressed electric field $\mathbf{E}_0$ polarized from left to right, the two nanoelements are indeed effectively in a parallel configuration, with the same potential difference between the two ends in each particle, and current flowing in the direction of the potential drop. In the quasi-series configuration of Fig. 3b, on the other hand, with applied electric field pointing from bottom to top, the displacement current flow (flux of the electric displacement vector) in the top nanoelement flows almost entirely into the second nanoelement, due to the induced potential difference between the two elements, and the current flow in the surrounding space is negligible.

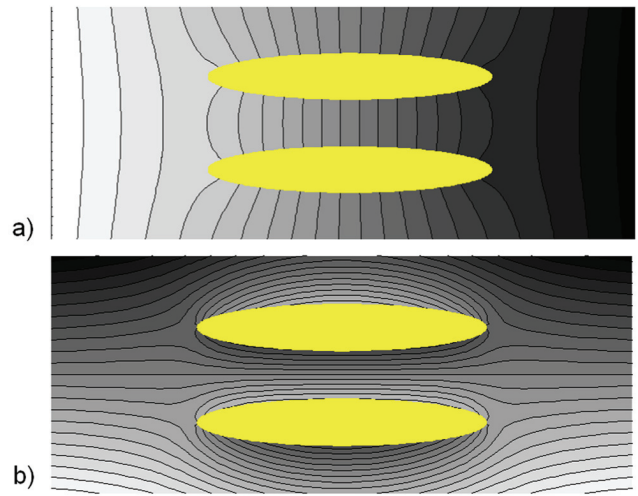

Fig. 3. (Color online). Potential distribution for a pair of "coupled" ellipsoids (yellow) with: a) $a_1 = a_2 = 60\,nm$, $b_1 = b_2 = 50\,nm$, $c_1 = c_2 = 5\,nm$, $d = 2(c_1 + c_2)$, $\varepsilon_1 = \varepsilon_2 = -\varepsilon_0$, $\mathbf{E_0} \perp \mathbf{d}$; b) $a_1 = a_2 = 60\,nm$, $b_1 = b_2 = 50\,nm$, $c_1 = c_2 = 5\,nm$, $d = 2(c_1 + c_2)$, $\varepsilon_1 = \varepsilon_2 = -\varepsilon_0$, $\mathbf{E_0} \parallel \mathbf{d}$. Lighter colors correspond to higher values of the potential.

In certain specific configurations the use of controlled (i.e., dependent) generators of Fig. 2 may not be necessary, and it may be avoided in the final circuit layout. For instance, in the case in which the nanoelements are effectively conjoined, as we have shown in [5]-[6], the nanoelements are effectively in series or parallel, depending on the orientation of the applied field as we discussed above, and the presence of controlled current generators would be redundant, since in these specific cases the two coupled circuits may be reduced to a single circuit with a series or parallel combination of the single impedances. We note, for instance, that in the parallel configuration the controlled current generators in the coupled circuit of Fig. 2 would be controlled by the voltage applied on their own terminals, since the two parallel elements see the same voltage difference applied to their terminals, resulting in the presence of one single effective parallel impedance. Analogous considerations hold for the series combination of nanocircuit elements.

It is clear, however, that when the distance between the pair of nanoelements increases with respect to their size, even though the orientation of the field may still satisfy the conditions for a quasi-series or quasi-parallel configuration, the field distribution may not be confined enough to make the nanoelements connected in the desired way. This effect is even more pronounced when multiple nanoelements are present and the design requires multiple interconnections among them. A generalization of the present theory to an arbitrary number of nanoparticles, and an arbitrary orientation of the impressed electric field with possibly varying its amplitude and direction for the different particle's position, is straightforward following this same analogy. This may model a complex nanocircuit board with many coupled circuit nanoelements. However, the design of the required circuit response may not be straightforward in the general case for the multiple coupling among the particles, which is much more complex than a standard circuit board, where the coupling among lumped elements happens only through their terminals and the wires connecting them.

We have been studying an extended mechanism for avoiding this unwanted coupling among nanocircuit elements that should not be "connected" on purpose. To this aim, it is possible to employ materials near their plasma frequency, with a permittivity near zero, to "insulate" the nanoelements from each other. Since the displacement current flowing in and out from a nanoelement is represented by the integral flux of the displacement current, owing to the continuity of the normal component of $\mathbf{D} = \varepsilon \mathbf{E}$, a material with zero permittivity (ε-near-zero, ENZ) surrounding a nanoelement would totally "block" (i.e., insolate) the current exchange of the element with the outside world. The anomalous properties of such ENZ materials fully confirm this expected behavior, as we have reported in [11]-[12].

On the other hand, in order to "connect" the terminals of two nanoelements that are not necessarily placed very close to each other, one can utilize materials with very large permittivity (ε-very-large, EVL). In this case, the flux of displacement current would be squeezed through suitable EVL channels, with a relatively low optical voltage drop, acting analogously to short circuits that interconnect lumped elements in a conventional circuit board. The use of ENZ and EVL materials for eliminating or inducing proper coupling among nanocircuit elements may help us in the design of complex nanocircuit setups. In [13] we underline the principles and utilities of such materials as optical nanoconnectors (EVL) and nanoinsulators (ENZ) in complex nanocircuit configurations.

## 4. Modeling the Effects of Substrates

The realization of a nanocircuit board may require the presence of a substrate over which the nanocircuit elements will be built and realized. The analysis of the present manuscript may also consider this situation as a coupling phenomenon between the nanoelement and its quasi-static image.

Consider a nanoparticle of permittivity $\varepsilon$ sitting over an infinite dielectric half-space of permittivity $\varepsilon_s$. The other half-space where the nanoparticle is present has permittivity $\varepsilon_0$, with the unit vector normal to the interface denoted by $\hat{\mathbf{n}}$ and the unit vector tangent to the interface denoted by $\hat{\mathbf{t}}$. A generic electric field $\mathbf{E_0}$ exciting the nanoparticle would induce a dipole moment $\mathbf{p_0} = \underline{\boldsymbol{\alpha}} \cdot \mathbf{E_0}$ on the particle itself, where $\underline{\boldsymbol{\alpha}}$ is the particle's polarizability tensor, which relates the induced electric dipole moment to the applied electric field. The effect of the presence of a substrate may be modeled in this quasi-static analysis as the presence of an image dipole symmetrically placed on the other side of the interface, as sketched in Fig. 4.

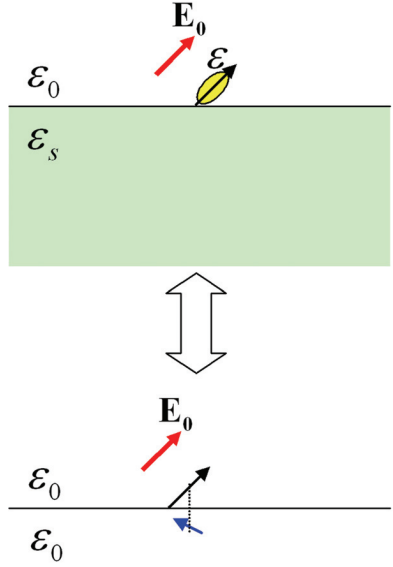

Fig. 4. (Color online). A nanocircuit element of permittivity $\varepsilon$ over a dielectric planar substrate of permittivity $\varepsilon_s$, formally equivalent to the coupling problem of two nanocircuit elements symmetrically displaced with respect to the interface, as consistent with the image theory.

The effective dipole moment **p** induced on the nanocircuit element is due to the sum of the impressed field $\mathbf{E}_0$ and the field from the image particle evaluated at the nanoelement:

$$\mathbf{p} = \underline{\boldsymbol{\alpha}} \cdot \left( \mathbf{E}_0 + \mathbf{E}_{image} \right). \tag{20}$$

The electric field $\mathbf{E}_{image}$ may be expressed in terms of the dipole moment induced on the image particle $\mathbf{p}_{image}$ as:

$$\mathbf{E}_{image}(\mathbf{r}) = \underline{\mathbf{G}}(\mathbf{r}) \cdot \mathbf{p}_{image}, \tag{21}$$

where $\overline{\mathbf{G}}(\mathbf{r})$ is the dyadic Green's function in free space [10].

The dipole moment of the image particle is given by [10]:

$$\mathbf{p}' = -\left(\frac{\varepsilon_2 - \varepsilon_1}{\varepsilon_2 + \varepsilon_1}\right)\left(\hat{\mathbf{t}}\hat{\mathbf{t}} - \hat{\mathbf{n}}\hat{\mathbf{n}}\right)\cdot \mathbf{p} = \bar{\mathbf{A}}\cdot \mathbf{p}, \tag{22}$$

leading to the following relations for the induced dipole moment on the nanocircuit element and on its image:

$$\begin{aligned}\mathbf{p} &= \left\{\bar{\boldsymbol{\alpha}}_{part} + \bar{\boldsymbol{\alpha}}_{part}\cdot \bar{\mathbf{G}}(2d)\cdot \left[\bar{\mathbf{I}} - \bar{\mathbf{A}}\cdot \bar{\boldsymbol{\alpha}}_{part}\cdot \bar{\mathbf{G}}(2d)\right]^{-1}\cdot \bar{\mathbf{A}}\cdot \bar{\boldsymbol{\alpha}}_{part}\right\}\cdot \mathbf{E}_{inc} \\ \mathbf{p}_{image} &= \left\{\left[\bar{\mathbf{I}} - \bar{\mathbf{A}}\cdot \bar{\boldsymbol{\alpha}}_{part}\cdot \bar{\mathbf{G}}(2d)\right]^{-1}\cdot \bar{\mathbf{A}}\cdot \bar{\boldsymbol{\alpha}}_{part}\right\}\cdot \mathbf{E}_{inc}\end{aligned}, \tag{23}$$

where $\underline{\mathbf{G}}(2d)$ is evaluated along the normal to the interface, connecting the nanoelement and its mage, at a distance $2d$ where $d$ is the distance of the nanoelement from the planar substrate.

These expressions formally provide the information about the coupling effect between a nanocircuit element and its planar substrate, highlighting how the problem is formally equivalent to the one of Section 3. The presence of a planar substrate may be modeled as an image nanocircuit element, as given by (22), and the equivalent coupled nanocircuit models presented in Section 3 may also hold here. We reiterate here that if these coupling phenomena between nanocircuit elements and their substrate are unwanted, it may be possible to avoid or drastically reduce them by utilizing ENZ and EVL materials as nanoinsulators and nanoconnectors, similar to what was suggested in Section 3. We discuss these issues in further details in [13].

## 5. *Conclusions*

In this contribution, we have modeled theoretically, in the framework of our nanocircuit paradigm for optical nanoparticles, the coupling between two electrically small nanoparticles immersed in a quasi-static optical electric field. We have found that the coupling affects the corresponding nanocircuit model by adding controlled (i.e., dependent) generators that depend on the optical voltages applied on the coupled particles. As a further contribution, we have also studied how the presence of a planar substrate placed underneath a nanocircuit element may affect its corresponding nanocircuit model, relating these issues to the coupling with a properly modeled image dipole. For reducing the unwanted coupling among nanocircuit elements, which may be more relevant than in standard circuit boards due to the different features of the problem, we suggest the use of nano-insulators and nano-connectors, respectively, in the form of ENZ and EVL materials. These are discussed in more detail elsewhere [13].

*Acknowledgements*

This work is supported in part by the U.S. Air Force Office of Scientific Research (AFOSR) grant number FA9550-05-1-0442. Andrea Alù was partially supported by the 2004 SUMMA Graduate Fellowship in Advanced Electromagnetics.